\documentstyle[12pt]{article}
\def\beq{\begin{equation}}
\def\eeq{\end{equation}}
\def\bea{\begin{eqnarray}}
\def\eea{\end{eqnarray}}
\def\nn{\nonumber}
\def\ba{\begin{array}}
\def\ea{\end{array}}
\def\d{\partial}
\def\v{\vert}
\def\l{\langle}
\def\r{\rangle}
\def\one{1\hskip -1mm{\rm l}}
\def\P{{\rm l}\hskip -.85mm{\rm P}}

\begin{document}
\rightline{TIFR/TH/97-01}
\rightline{January 1997}
\baselineskip16pt
\smallskip
\begin{center}
{\large \bf \sf
Multi-parameter deformed and nonstandard $Y(gl_M)$ Yangian symmetry \\
in a novel class of spin Calogero-Sutherland models }\\

\vspace{1.5cm}

{\sf B. Basu-Mallick$^1$\footnote{
E-mail address: biru@theory.tifr.res.in.
Address from 1st May, 1997:
Department of Physics, Faculty of Science, University of Tokyo,
Hongo 7-3-1, Bunkyo-ku, Tokyo 113 }
and Anjan Kundu$^2$\footnote{E-mail address:
anjan@tnp.saha.ernet.in } },

\bigskip

{\em $^1$Theoretical Physics Group, \\
Tata Institute of Fundamental Research, \\
Homi Bhabha Road, Mumbai-400 005, India} \\

\bigskip

{\em $^2$Theory Group, \\
Saha Institute of Nuclear Physics, \\
1/AF Bidhan Nagar, Calcutta 700 064, India } \\
\bigskip

\end{center}

\vspace {1.25 cm}
\baselineskip=20pt
\noindent {\bf Abstract }

It is well known through a recent work of Bernard, Gaudin, Haldane and
Pasquier (BGHP) that the usual spin Calogero-Sutherland (CS) model,
containing particles with $M$ internal degrees of freedom, respects the
$Y(gl_M)$ Yangian symmetry. By following and suitably modifying the approach
of BGHP, in this article we construct a novel class of spin CS models
which exhibit multi-parameter deformed or `nonstandard' variants of
$Y(gl_M)$ Yangian symmetry. An interesting feature
of such CS Hamiltonians is that they
contain many-body  spin dependent interactions, which can be calculated
directly from the associated rational solutions of Yang-Baxter equation.
Moreover, these spin dependent interactions often lead to  `anyon like'
representations of permutation algebra on the combined internal space of
all particles. We also find out the general forms of conserved quantities
as well as Lax pairs for the above mentioned class of spin CS models,
and describe the method of constructing their exact wave functions.

\vspace {.75 cm}

\vspace {.05 cm}

\newpage

\baselineskip=22pt
\noindent \section {Introduction }
\renewcommand{\theequation}{1.{\arabic{equation}}}
\setcounter{equation}{0}

\medskip
Algebraic structures of (1+1) dimensional quantum
integrable systems with long ranged interactions and their
close connection with diverse subjects like  conformal field theory,
matrix models, fractional statistics, quantum Hall effect etc. have
attracted intense attention in recent years [1-17].
In particular it is found that,  commutation relations between the
conserved quantities of well known spin Calogero-Sutherland (CS)
model, given by the Hamiltonian
\beq
H ~=~  -{1\over 2} \sum_{i=1}^N ~ \left ( { \d \over \d x_i }
\right )^2 ~+
{ \pi^2 \over L^2 }~ \sum_{i<j}~{   \beta ( \beta + P_{ij}  )  \over
\sin^2 { \pi \over L} (x_i -x_j)  }~,
\label {a1}
\eeq
where $\beta $ is a coupling constant and
$P_{ij}$ is the permutation operator interchanging the `spins'
of $i$-th and $j$-th particles, generate the  $Y(gl_M)$
Yangian algebra [3]. This  $Y(gl_M)$ Yangian algebra [18,19]
can be defined through the operator valued elements of a
$M\times M$ dimensional monodromy matrix $T^0(u)$,
which obeys the quantum Yang-Baxter equation (QYBE)
\beq
R_{00'}(u-v) \left ( T^0(u) \otimes \one \right )
 \left ( \one \otimes  T^{0'}(v) \right ) ~=~
 \left ( \one \otimes  T^{0'}(v) \right )
 \left ( T^0(u) \otimes \one \right )  R_{00'}(u-v) \, .
\label {a2}
\eeq
Here $u$ and $v$ are  spectral parameters and
 the $M^2 \times M^2$ dimensional rational $R(u-v)$ matrix,  having usual
$c$-number valued elements, is taken as
\beq
R_{00'} (u-v)  ~=~   (u-v) \, \one  \, + \, \beta \,  P_{00'} .
\label {a3}
\eeq
So the conserved quantities of spin CS  model (\ref {a1})
yield a realisation of
$T^0(u)$ matrix satisfying this QYBE (\ref {a2}).
Moreover, the spin CS Hamiltonian (\ref {a1}) can be reproduced in
a simple way from the quantum determinant associated with such
monodromy matrix. This close connection between $Y(gl_M)$
Yangian algebra and spin CS model (\ref {a1}) helps to
find out the related orthogonal basis of eigenvectors and
might also play an important role in calculating
various dynamical correlation functions [20].

However, it is worth noting that there exist a class of
rational $R$ matrices  which satisfy the Yang-Baxter equation (YBE)
\beq
  R_{00'} (u-v) \, R_{00''} (u-w) \, R_{0'0''} (v-w)  ~=~
  R_{0'0''} (v-w) \, R_{00''} (u-w) \, R_{00'} (u-v) \,  ,
\label {a4}
\eeq
(here a matrix like $R_{ab}(u-v)$ acts nontrivially only on the $a$-th
and $b$-th vector spaces)
and reduce to the $R$ matrix (\ref {a3}) at some particular limits
of related deformation parameters.
These generalisations of rational solution (\ref {a3}) are interestingly
connected with  various multi-parameter dependent deformations
of $Y(gl_M)$ Yangian algebra [21-24] and some integrable lattice models
with local interactions [25-27].
The general form of such rational solutions,
as well as their `nonstandard' variants (which will be explained
shortly), might be written as
\beq
R_{00'} (u-v)  ~=~   (u-v) \, Q_{00'} \, + \, \beta \, P_{00'},
\label {a5}
\eeq
where $P_{00'}$ is the usual permutation matrix which interchanges
two vectors associated with $0$-th and $0'$-th auxiliary spaces,
 and $Q_{00'}$ is another $M^2 \times M^2$ dimensional
matrix whose elements may depend on  deformation parameters.
By substituting (\ref {a5}) to (\ref {a4})
and using the above mentioned property of $P_{00'}$,
it is  easy to check that the $R$ matrix (\ref {a5})
would be a valid solution of YBE, provided
the corresponding $Q$ matrix satisfies only two conditions:
\beq
Q_{00'} \, Q_{00''} \, Q_{0'0''} ~=~ Q_{0'0''} \, Q_{00''} \, Q_{00'}
\, , ~~~~~ Q_{00'} \, Q_{0'0}  ~=~ \one \, .
\label {a6}
\eeq
Thus, any solution of eqn.(\ref {a6})
will give us a rational $R$ matrix in the form (\ref {a5}),
which, in turn, can be  inserted to QYBE (\ref {a2}) for obtaining
a possible extension of $Y(gl_M)$ Yangian algebra.
The simplest solution of eqn.(\ref {a6})
is evidently given by $Q_{00'} = \one $, which reproduces the original
$R$ matrix (\ref {a3}) and the standard $Y(gl_M)$ Yangian algebra.
However, in general, a solution of eqn.(\ref {a6}) might also
 depend on a set of continuous deformation parameters like $\{h_p\}$. So
these parameters would naturally appear in the defining relations of
corresponding extended Yangian algebra.
Moreover, the solutions of eqn.(\ref {a6})
often admit a Taylor series expansion
in the form (up to an over all normalisation factor)
\beq
Q_{00'} ~=~ \one ~+~ \sum_p \, h_p \, Q^p_{00'} ~+~ \sum_{p,q} \,
h_p h_q \, Q^{pq}_{00'} ~+~ \cdots \, ,
\label {a7}
\eeq
where the leading term is an identity operator.
Consequently the multi-parameter dependent Yangian algebras,
generated through such $Q$ matrices,
would reduce to  standard $Y(gl_M)$
algebra at the limit $h_i \rightarrow 0 $ for all $i$. Though many
mathematical properties of these  multi-parameter
deformed $Y(gl_M)$ Yangians have been
studied earlier, the important problem of constructing quantum
integrable models with long range interactions which would respect such
Yangian symmetries has received little attention till now. So,
it should be quite encouraging to enquire whether
there exist some new variant of  spin CS Hamiltonian (\ref {a1})
which would exhibit a deformed $Y(gl_M)$ Yangian symmetry associated
with rational $R$ matrix (\ref {a5}).

Furthermore, one may like to seek answer of the above
mentioned problem in a slightly different context when the
$Q$ matrix, which is obtained as a  solution of eqn.(\ref {a6}),
can not be expanded in the form (\ref {a7}).
Though in many previous works [21-24] only the type of
$Q$ matrices that can be expanded as
(\ref {a7}) were discussed, in this article we  construct some other 
forms of 
$Q$ matrices which do not yield identity operator as the $0$-th
order term in their power series expansion.
Consequently the Yangian algebras, generated through such
`nonstandard' $Q$ matrices and corresponding rational solutions
(\ref {a5}), will not reduce to  $Y(gl_M)$ algebra at
the limit $h_i \rightarrow 0$.  Due to this reason,
those Yangians may be called as  `nonstandard' variants
of $Y(gl_M)$ Yangian algebra.

In this article our main aim is to
develope a rather general framework for
constructing a large class of quantum integrable spin CS Hamiltonians,
each of which would exhibit an extended
(i.e., multi-parameter deformed or `nonstandard' variants of)
$Y(gl_M)$ Yangian symmetry. So in sec.2 we start with the
rational $R$ matrix (\ref {a5}), but do not assume any particular form
of corresponding $Q$ matrix, and
attempt to construct a spin CS Hamiltonian from the quantum determinant
like object of related Yangian algebra. To this end,
we closely follow and suitably modify the pioneering approach of ref.3,
where a realisation of $Y(gl_M)$ algebra is obtained through the conserved
quantities of usual spin CS Hamiltonian (\ref {a1}). Subsequently,
we also describe the method of constructing exact wave functions for
spin CS models which exhibit the extended
$Y(gl_M)$ Yangian symmetries.  Next, in sec.3,
we examine the question of quantum integrability for the above mentioned
class of spin CS models and write down the general forms of
their Lax pairs as well as conserved quantities.
Finally, in sec.4, we consider some specific
examples of $Q$ matrices which  satisfy the
conditions (\ref {a6}),  and attempt to find out the concrete forms of
related spin CS Hamiltonians,  conserved
quantities and Lax pairs. Sec.5 is the concluding section.

\vspace{1cm}

\noindent \section { Construction of spin CS Hamiltonian with
extended $Y(gl_M)$ Yangian symmetry }
\renewcommand{\theequation}{2.{\arabic{equation}}}
\setcounter{equation}{0}

\medskip

Here we like to find out the general form of a
spin CS Hamiltonian, whose conserved quantities would produce
a realisation of extended $Y(gl_M)$ Yangian algebra associated with
the $R$ matrix (\ref {a5}).
So, in our discussion in this section, we shall not assume
any specific form of the corresponding $Q$ matrix and only use
the fact that it satisfies the two conditions (\ref {a6}).

However, for our purpose of
constructing the above mentioned spin CS Hamiltonian,
it would be convenient to briefly recall the method of
generating the monodromy matrix for quantum
integrable spin chains which contain only local interactions [28-30].
To obtain the monodromy matrix for such a spin chain,
one considers a Lax operator $L^0_i(u)$ whose
matrix elements (operator valued) depend  only
on the spin variables of $i$-th lattice site
and satisfy the QYBE given by
\beq
R_{00'}(u-v) \left ( L^0_i(u) \otimes \one \right )
 \left ( \one \otimes  L^{0'}_i(v) \right ) ~=~
 \left ( \one \otimes  L^{0'}_i(v) \right )
 \left ( L^0_i(u) \otimes \one \right )  R_{00'}(u-v) \, ,
\label {b1}
\eeq
where
$R_{00'}(u-v)$ is a solution of YBE (\ref {a4}). In a similar way,
one can associate  a Lax operator on every lattice site of the spin
chain. The monodromy matrix for this spin chain,
containing $N$ number of lattice sites,
is generated by multiplying all these Lax operators on the auxilliary
space as
\beq
 T^0(u) ~=~ L^0_N(u)  L^0_{N-1}(u)\cdots  L^0_i(u) \cdots  L^0_1(u) \, .
\label {b2}
\eeq
By applying (\ref {b1}) and  also using
the fact that the spin variables at different lattice sites are
commuting operators, it is easy to prove that the monodromy matrix
(\ref {b2}) would also satisfy  QYBE (\ref {a2})
for the same $R_{00'}(u-v)$ matrix appearing in (\ref {b1}).
So, by multiplying some  `local' solutions of QYBE,  one can
also generate its `global' solution.

Now,
for  finding out the Lax operator of a spin chain
associated with the rational solution
(\ref {a5}), we follow the standard procedure of treating
the second auxiliary space in
$R_{0i}(u-\eta_i)=L^0_{i}(u)$ matrix as a `quantum' space and write down the
corresponding $L^0_{i}(u) $ operator as
\beq
L^0_{i}(u) ~=~ Q_{0i} ~+~ \beta \, { P_{0i} \over { u - \eta_i } } \, \, ,
\label {b3}
\eeq
where $\eta_i$ is an arbitrary constant.
By using eqn.(\ref {a6}), one can also directly check that the Lax operator
(\ref {b3}) and $R$ matrix (\ref {a5}) satisfy the  QYBE (\ref {b1}).
However the Lax operator (\ref {b3}) is not a good choice for our present
purpose, since its elements do not  contain yet any coordinate or momentum
variable which can be related to some new type of spin CS model.
So we modify this Lax operator in the following way:
\beq
{\hat L}^0_{i}(u) ~=~
Q_{0i} ~+~ \beta \, { P_{0i} \over { u - {\hat D}_i } } \, \, ,
\label {b4}
\eeq
where ${\hat D}_i$s $(i \in [1,N])$, the so called Dunkl operators,
are defined as [3,31]
\beq
{\hat D}_i ~=~ z_i { \d \over \d z_i } ~+~
 \beta \, \sum_{j>i} \, \theta_{ij} K_{ij}
~-~ \beta \, \sum_{j<i } \theta_{ji} K_{ij} \,  ,
\label {b5}
\eeq
 $ z_i = e^{{ 2\pi i \over L} x_i} $, $ \theta_{ij} =
{ z_i \over z_i - z_j } $,  and $K_{ij}$'s are the coordinate
exchange operators which obey the relations
\bea
&~~ \quad \quad  K_{ij} z_i ~=~ z_j K_{ij},~~K_{ij} { \d \over \d z_i } ~=~
{ \d \over \d z_j } K_{ij}, ~~K_{ij} z_l ~=~ z_l K_{ij}~,~ \nn &
\quad \quad  \quad   \quad  (2.6a)  \\
&~~ \quad \quad  K_{ij}^2 ~=~ 1, ~~~K_{ij}K_{jl} ~=~ K_{il}K_{ij} ~=~
K_{jl} K_{il} ~, ~~~[\, K_{ij}, K_{lm} \, ] ~=~ 0~,~ \nn  &
\quad  \quad \quad  \quad  (2.6b)
\eea
\addtocounter{equation}{1}
\noindent $ i, ~j,~l,~m ~$ being all different indices.
The Lax operator (\ref {b4}) may now be related to the
$i$-th particle, rather than the $i$-th lattice site,
which  moves continuously on a circle (we have assumed an ordering
among the particles). Since the Dunkl operators
 (\ref {b5}) do not act on the spin
degrees of freedom, it is evident that this new Lax operator
would  also satisfy the QYBE (\ref {b1}) for our choice of rational
$R$-matrix (\ref {a5}).   Moreover, by using (\ref {b5}) and
(2.6), one can check that these
Dunkl operators satisfy the standard commutation relations
\bea
&~~~~~~~~~~ \left [ \, {\hat D}_i \, , \, {\hat D}_j \, \right ] \, =
\, 0 \, , ~~~~~
 \left [ \, K_{i,i+1} \, , \, {\hat D}_k \, \right ] \, = \, 0 \, ,\nn
~~~~~~~~~~~~~~~~~~&(2.7a,b) \\
&~~~~~~~~~~K_{i,i+1} {\hat D}_i  \, -  \, {\hat D}_{i+1} K_{i,i+1} \, =
\,  \beta \, , ~~~~
\left [ \, K_{i,i+1} \, , \, \Delta_N(u)   \, \right ] \, = \, 0 \, ,\nn
~~~~~~~~~~~~~~~~~~&(2.7c,d)
\eea
\addtocounter{equation}{1}
where $k \neq i,\, i+1 $ and
$\, \Delta_N(u) ~=~ \prod_{i=1}^N \left ( u - {\hat D}_i \right ) $.
Now, by applying the relation (2.7a), it is easy to see
that the matrix elements of $ {\hat L}^0_i(u) $ would commute with
that of ${\hat L}^0_j(u)$,  when $i \neq j$. Consequently, by using
(\ref {b2}), we can construct a monodromy matrix like
\bea
{\hat T}^0(u) ~&=~
\left ( Q_{0N} + \beta { P_{0N} \over u - {\hat D}_N } \right )
\left ( Q_{0,N-1} + \beta { P_{0,N-1} \over u - {\hat D}_{N-1} } \right )
\cdots \left ( Q_{01} + \beta { P_{01} \over u - {\hat D}_1 } \right ) \nn
\\
&= \left  \{ \Delta_N(u) \right \}^{-1} \, \prod_{i=N}^1 \,
\left [\, (u - {\hat D}_i ) Q_{0i} + \beta P_{0i}  \, \right ]  \, ,
~~~~~~~~~~~~~~~~~~~~
\label {b8}
\eea
which would satisfy the QYBE (\ref {a2}) and, therefore, yield
a realisation of extended $Y(gl_M)$ Yangian algebra
associated with the rational $R$-matrix (\ref {a5}).

However, the above constructed
monodromy matrix still contains the coordinate exchange operators
$K_{ij}$ which we want to eliminate from our final expression.
So we define a projection operator as
\beq
\Pi^* (K_{ij}) ~=~ {\tilde P}_{ij} \, ,
\label {b9}
\eeq
where
$ {\tilde P}_{ij}$s are some yet undetermined spin dependent operators
which would act on the combined
internal space of all particles (i.e., on $ {\cal F} ~\equiv ~
\underbrace { C^M \otimes C^M \otimes  \cdots \otimes C^M }_N ~$).
The projection operator in eqn.(\ref {b9}) is defined
in the sense that one should replace $K_{ij} $ by
$ {\tilde P}_{ij} $, only after moving
$K_{ij} $ in the extreme right side of an expression.
However,  it is expected that $\Pi^*$
will produce the same result while
acting on the l.h.s. and r.h.s. of each equation appearing in (2.6b).
By using such consistency conditions,
  it is easy to prove that ${\tilde P}_{ij}$s
must yield a representation of following permutation algebra
on the space ${\cal F}$:
\beq
  {\cal P}_{ij}^2 ~=~ 1,
~~~{\cal P}_{ij}{\cal P}_{jl} ~=~ {\cal P}_{il}{\cal P}_{ij} ~=~
{\cal P}_{jl} {\cal P}_{il} ~, ~~~[ {\cal P}_{ij},
{\cal P}_{lm} ] ~=~ 0~ ,
\label {b10}
\eeq
$ i, ~j,~l,~m $  being all different indices. As it is well known,
the above permutation algebra  can be  generated by
the `nearest neighbour' transposition elements
${\cal P}_{i,i+1}$ $(i \in [1,N-1] )$,
which satisfy the relations
\bea
&& {\cal P}_{i,i+1} {\cal P}_{i+1,i+2} {\cal P}_{i,i+1} \,  = \,
{\cal P}_{i+1,i+2} {\cal P}_{i,i+1} {\cal P}_{i+1,i+2} \, ,~
\left [  {\cal P}_{i, i+1 },{\cal P}_{k, k+1} \right ] \, = \, 0\, , ~
 {\cal P}_{i,i+1}^2 \, = \,  \one \, ,~~~~~~~~ \nn\\
&& \hskip 13.15 cm \nn  (2.11a,b,c)
\eea
\addtocounter{equation}{1}
where  $\v i-k \v > 1 $. All  other `non-nearest neighbour' transposition
elements like  ${\cal P}_{ij} $ (with $j-i >1 $) can be expressed
through these generators as
\beq
{\cal P}_{ij} ~~=~~ \left (
{\cal P}_{i,i+1} {\cal P}_{i+1,i+2} \cdots  {\cal P}_{j-2,j-1} \right )
\, {\cal P}_{j-1,j} \, \left ( {\cal P}_{j-2,j-1} \cdots
{\cal P}_{i+1,i+2} {\cal P}_{i,i+1} \right ) ~.
\label {b12}
\eeq
So, the projection operator $\Pi^*$ will be completely defined
if we specify its action only on $N-1$ number of
coordinate exchange operators like $K_{i,i+1}$.

In this context one may note that, while
constructing a realisation of $Y(gl_M)$ Yangian algebra
through the conserved quantities of usual spin CS model (\ref {a1}),
a projection operator is defined in ref.3 as:  $\Pi (K_{ij}) =  P_{ij}.$
Here $P_{ij}$ is the standard permutation operator which acts
on the space ${\cal F}$ as
\beq
 P_{ij} ~\v \alpha_1 \alpha_2  \, \cdots  \, \alpha_i \, \cdots \, \alpha_j
 \, \cdots \, \alpha_N \r ~=~
 \v \alpha_1 \alpha_2   \, \cdots  \, \alpha_j  \, \cdots  \,  \alpha_i
 \, \cdots  \, \alpha_N \r ~,
\label {b13}
\eeq
where $ \v \alpha_1 \alpha_2   \cdots \alpha_i \cdots \alpha_N \r $
(with  $\alpha_i  \, \in [1,M] $)  represents  a particular
spin configuration of  $N$ particles.   It is obvious
that  this $P_{ij}$  produces  a representation of the
permutation algebra (\ref {b10}).
However it is already known that,
by taking appropriate limits of some braid group representations
which also satisfy the Hecke algebra,
one can easily construst many other inequivalent
representations of permutation algebra  (\ref {b10})
on the space ${\cal F}$ [32]. So, while defining a projection
operator in eqn.(\ref {b9}),  we have not chosen
 ${\tilde P}_{ij} = P_{ij}$ from the very beginning. In fact,
our aim in the following is to find out the precise form of this
 ${\tilde P}_{ij}$, by demanding that the projection of
monodromy matrix (\ref {b8}), i.e.
\beq
T^0(u) ~=~ \Pi^* \left [ {\hat T}^0(u) \right ] \, ,
\label {b14}
\eeq
would also satisfy QYBE (\ref {a2}), when the corresponding
$R$ matrix is taken as (\ref {a5}). Evidently, this
$T^0(u)$ would give a solution of QYBE
if $ {\hat T}^0(u)$ satisfies the condition:
$
 \Pi^* \left [ {\hat T}^0(u) {\hat T}^0(v) \right ] ~=~ $$
 \Pi^* \left [ {\hat T}^0(u)  \right ]
 \Pi^* \left [ {\hat T}^0(v) \right ] $, or, equivalently
\beq
 \Pi^* \left [ K_{i,i+1} {\tilde P}_{i,i+1} {\hat T}^0(u) \right ]
~=~ \Pi^* \left [ {\hat T}^0(u) \right ] \, .
\label {b15}
\eeq
By inserting the explicit form of $ {\hat T}^0(u)$
 (\ref {b8}) to the above equation and assuming that
${\tilde P}_{i,i+1}$ acts nontrivially only on $i$-th
and $(i+1)$-th spin spaces,
(\ref {b15}) can be simplified as
\bea
& \Pi^* \left [ \, K_{i,i+1} \,  \left ( \, (u -{\hat D}_i) Q_{0i}
+ \beta P_{0i} \, \right )
\left ( \, (u -{\hat D}_{i+1}) Q_{0,i+1} + \beta P_{0,i+1} \, \right )
\, \right ] ~~~~~~~~~ \nn
\\
&~~~~~~~~~~~~=~ {\tilde P}_{i,i+1} \,
 \Pi^* \left [ \,  \left ( \, (u -{\hat D}_i) Q_{0i}
+ \beta P_{0i} \, \right )
\left ( \, (u -{\hat D}_{i+1}) Q_{0,i+1} + \beta P_{0,i+1} \, \right )
 \, \right ]  .~~~
\label {b16}
\eea
Furthermore,  by using  eqns.(2.7) and (\ref {b9}), the
condition (\ref {b16}) can be finally expressed as
\beq
{\cal A}_1 \, \Pi^* \left [ \, ( u -{\hat D}_i)
( u -{\hat D}_{i+1} ) \, \right ]
\, + \, \beta \,  {\cal A}_2  \,  \Pi^*  ( u -{\hat D}_i)
\, + \, \beta \,  {\cal A}_3  \,  \Pi^*  ( u -{\hat D}_{i+1}) \, + \,
\beta^2 {\cal A}_4 \,=~0 \, ,
\label {b17}
\eeq
where
\bea
&{\cal A}_1
= \left [ \, Q_{0i} Q_{0,i+1} , {\tilde P}_{i,i+1} \, \right ]\, ,
~~~{\cal A}_2 = P_{0i} Q_{0,i+1} {\tilde P}_{i,i+1} - {\tilde P}_{i,i+1}
Q_{0i} P_{0,i+1} \, , ~~~~~~\nn \\
&{\cal A}_3 = Q_{0i} P_{0,i+1} {\tilde P}_{i,i+1} - {\tilde P}_{i,i+1}
P_{0i} Q_{0,i+1}  ,~
{\cal A}_4 = \left [ P_{0i} P_{0,i+1}  , {\tilde P}_{i,i+1} \right ] +
 P_{0i} Q_{0,i+1} - Q_{0i} P_{0,i+1} \, .~~~~~~ \nn
\eea
Now, it is immensely interesting to observe that we can set
${\cal A}_1 = {\cal A}_2 = {\cal A}_3 = {\cal A}_4 = 0 $,
provided we assume the simple relation
\beq
\Pi^* ( K_{i,i+1} ) ~=~
{\tilde P}_{i,i+1} ~=~ Q_{i,i+1} P_{i,i+1} \, ,
\label {b18}
\eeq
and also use the two general conditions (\ref {a6}) satisfied by the $Q$
matrix. Consequently, the projected monodromy
 matrix (\ref {b14}) would give us a novel realisation
of extended $Y(gl_M)$ Yangian algebra, if the spin dependent operator
${\tilde P}_{i,i+1}$ occuring in the
 relation (\ref {b9}) is defined according to eqn.(\ref {b18}).
It is worth noting that for the special case
$Q_{i,i+1} = \one $ one gets back
$ \Pi^* (K_{i,i+1}) \, = \,  P_{i,i+1}$, which was used in ref.3 to
find out a realisation of $Y(gl_M)$ Yangian algebra
through the conserved quantities of standard spin CS model (\ref {a1}).
Now, by applying again the conditions (\ref {a6}) and
well known properties of $P_{i,i+1}$, it is easy to
verify that  ${\tilde P}_{i, i+1}$ operators defined by (\ref {b18})
indeed satisfy the permutation algebra (2.11). So,
the form of a general ${\tilde P}_{ij}$ (with $j-i > 1 $)
can be obtained by using (\ref {b12})
and (\ref {b18}) as
\beq
\Pi^* \left ( K_{ij} \right ) ~=~
{\tilde P}_{ij}~=~ \left(\, Q_{i,i+1}  Q_{i,i+2} \, \cdots  \, Q_{ij}
\, \right) \, P_{ij} \,
\left( \, Q_{j-1,i} Q_{j-2,i} \, \cdots  \, Q_{i+1,i} \, \right) \, .
\label {b19}
\eeq
Thus our relations (\ref {b18}) and (\ref {b19}) give
a general prescription of defining the projection operation $\Pi^*$,
which can be used to construct the realisation (\ref {b14}) of extended
$Y(gl_M)$ algebra
associated with any possible solution of YBE
written in the form (\ref {a5}).

Next, we try to find out the spin CS
Hamiltonian which would exhibit the extended $Y(gl_M)$ Yangian
symmetry and, therefore, commute with all elements of the
$T^0(u)$ matrix (\ref {b14}). For the special
case $Q_{0,0'} = \one $, it is possible to derive  such a
spin CS Hamiltonian (\ref {a1}) from the quantum
determinant of $Y(gl_M)$ algebra [3]. However,
it is difficult to obtain the quantum determinant of
extended $Y(gl_M)$ Yangian algebra associated with a
rational solution (\ref {a5}), unless we take some
specific form of the corresponding $Q$ matrix. So,
we will describe here a rather adhoc procedure of constructing
quantum determinant like objects,
which would commute with all elements of
$T^0(u)$ (\ref {b14}) for any given choice of the related
$Q$ matrix.  For this purpose we first define
a set of operators $I_n$ through the
power series expansion:
$ \prod_{i=1}^N \left ( u - {\hat D}_i \right ) =
\sum_{n=0}^N  I_n \, u^{N-n} $ and
use  eqn.(2.7) to find that
\beq
 \left [ \,  f (I_1, I_2, \cdots , I_N) \, ,
\,  {\hat T}^0(u) \, \right ] ~=~0 \, , ~~~~~
\left [ \, f (I_1, I_2, \cdots , I_N) \, , \,
{\tilde P}_{i,i+1} K_{i,i+1} \, \right ] ~=~0 \, ,
\label {b20}
\eeq
where $f (I_1, I_2, \cdots , I_N)$
is an arbitrary polynomial function
of $I_n$s and ${\hat T}^0(u)$ is given by (\ref {b8}). Now, by applying
(\ref {b20}) and (\ref {b15}),  it is easy to see that 
$\Pi^* \left [ f (I_1, I_2, \cdots , I_N) \right ] $ will 
commute with all elements of the projected monodromy matrix
(\ref {b14}). In particular, the choice
$ {2 \pi^2 \over L^2 } \Pi^* \left( I_1^2 - I_2 \right) =
{2 \pi^2 \over L^2 }
\Pi^* \left ( \sum_{i=1}^N \, {\hat D}_i^2  \right ) $ would give us such
a Casimir operator, which can be written more explicitly by using
eqn.(\ref {b5}) as
\beq
{\tilde H} ~=~ {2 \pi^2 \over L^2 } \,
\Pi^* \left ( \sum_{i=1}^N \, {\hat D}_i^2  \right )
~=~  -{1\over 2} \sum_{i=1}^N ~ \left ( { \d \over \d x_i }
\right )^2 ~+
{ \pi^2 \over L^2 }~ \sum_{i<j}~{   \beta ( \beta + {\tilde P}_{ij}  )
 \over \sin^2 { \pi \over L} (x_i -x_j)  }~.
\label {b21}
\eeq
Evidently the above expression can be interpreted as a
spin CS Hamiltonian,
where the operators ${\tilde P}_{ij}$, defined by eqns.(\ref {b18}) and
(\ref {b19}), produce the spin dependent interactions.
It is obvious that for the special case
$Q = \one $,  (\ref {b21}) reduces to the original
spin CS Hamiltonian (\ref {a1})  containing  only two-body 
spin dependent interactions.
However, it is clear from eqn.(\ref {b19}) that,
${\tilde P}_{ij}$ would generally depend on all
spin variables associated with $j-i+1$ number of
particles indexed by $i, \, i+1, \,  \cdots , \,  j $.
So in contrast to the case of usual permutation
operator $P_{ij}$, which acts nontrivially only on the spin spaces
of $i$-th and $j$-th particles and
represents a two-body interaction, the  new operator
${\tilde P}_{ij}$ would lead to a many-body  spin dependent
interaction in the Hamiltonian (\ref {b21}).

It should be mentioned that a spin CS Hamiltonian like (\ref {b21})
was studied earlier and solved exactly by applying a `generalised'
antisymmetric projection operator on the eigenfunctions of
Dunkl operators [32]. However, in contrast to the present case,
no method was prescribed in ref.32
about the way of constructing spin dependent operators
${\tilde P}_{ij}$ from a given solution of YBE.
So our analysis not only reveals a rich symmetry structure
of the Hamiltonian (\ref {b21}),
but also prescribes a very convenient method of constructing
such Hamiltonian through the rational solution
of YBE (\ref {a5}). In the following, we like to
briefly recall the procedure of solving the Hamiltonian
(\ref {b21}) and show that the rational solution
(\ref {a5}) also plays a crucial role in finding out
the corresponding wave functions.  So we make an ansatz
for the wave function ${\tilde \psi}$ of spin CS
Hamiltonian (\ref {b21}) as [3,32]:
 $
 {\tilde \psi } ( x_1 , \cdots  ,  x_N ; \, \alpha_1 , \cdots  ,
 \alpha_N ) \, = \, \left [ \prod_{i<j} \sin { \pi \over L} (x_i -x_j )
 \right ]^\beta \, {\tilde \phi } ( x_1 , \cdots , x_N ; \,
 \alpha_1, \cdots  ,  \alpha_N ) \, , $
where it is assumed that $\beta >0$ to avoid singularity
at the origin.
Now, by applying the canonical commutation relations
$ [ { \d\over \d x_j } , x_k ] = \delta_{jk} $,
one may easily find that
\beq
 {\tilde H} {\tilde \psi } ~=~
 {2\pi^2 \over L^2 } \,
  \left [ \prod_{i<j} \sin { \pi \over L} (x_i -x_j )
\right ]^\beta \, \Pi_{(-)}^* \left ( {\cal H}^* \right ) \,
{\tilde \phi } \, ,
\label {b22}
\eeq
where
 $ \Pi_{(-)}^* $ is a projection operator defined by
 $ \Pi_{(-)}^* \left(K_{ij}\right) ~=~ - \, {\tilde P}_{ij} $,
 $ \,  {\cal H}^*  ~=~ \sum_{i=1}^N \, d_i^2 $ and
\beq
d_i ~=~ z_i { \d \over \d z_i } ~+~
\beta \left (~ i - { N+1 \over 2 } ~\right )
~-~ \beta ~\sum_{j>i }  \theta_{ij} \left ( K_{ij} -1 \right )
~+~ \beta ~\sum_{j<i } \theta_{ji}  \left ( K_{ij} -1 \right )
 \, .
\label{b23}
\eeq
These $d_i$s may be considered as a `gauge
transformed' variant of the Dunkl operators ${\hat D}_i$
(\ref {b5}) and they satisfy an algebra quite similar to (2.7).
So  one can construct the eigenvectors
of ${\cal H}^*$ by simultaneously diagonalising the
mutually commuting set of
operators $d_i$.  To this end, however,
it is helpful to make an ordering [3]  of the corresponding
basis elements characterised by the monomials like
$ \,  z_1^{\lambda_1 }
z_2^{\lambda_2 } \cdots z_N^{\lambda }  \, , $
where $\{ \lambda_1 , \cdots , \lambda_N \} \equiv  [\lambda ] $
is a sequence of non-negative integers
with  homogeneity  $ \lambda = \sum_{i=1}^N \lambda_i $.
Due to such ordering of monomials
within a given  homogeneity sector,
it turns out that the  operators $d_i$ and ${\cal H}^*$ can be represented
through some simple block-triangular matrices.
By taking advantage of this block-triangular property, it is not
difficult to find that
\beq
{\cal H}^* ~\xi_{[\lambda ]} (z_1,z_2, \cdots ,z_N) ~=~
 \sum_{i=1}^N ~
\left [ \lambda'_i - \beta \left ( { N+1 \over 2 } -i \right ) \right ]^2 ~
~\xi_{[\lambda ]} (z_1,z_2, \cdots ,z_N)  \, ,
\label {b24}
\eeq
where  $[\lambda' ] $ is a permutation of sequence $ [\lambda ]$
with the property
$ \lambda'_1 \leq \lambda'_2 \leq \cdots  \leq \lambda'_N $, and
the asymmetric Jack polynomial
$~\xi_{[\lambda ]} (z_1,z_2, \cdots ,z_N) $ is a suitable
linear combination of $ \,  z_1^{\lambda_1 }
z_2^{\lambda_2 } \cdots z_N^{\lambda }  \, , $
and other monomials of relatively lower orders.
Though it is rather
difficult to write down the general form of this
$\xi_{[\lambda ]} $, one can derive it easily for the case of
low-lying excitations through diagonalisation of small block-triangular
matrices. Furthermore, by using any given eigenfunction of ${\cal H}^* $,
it is possible to construct a set of degenerate
wave functions  corresponding to the spin CS Hamiltonian
(\ref {b21}) in the following way. Let
$\rho (\alpha_1 , \alpha_2 , \cdots , \alpha_N ) $ be an arbitrary
spin dependent function and
${\tilde \Lambda } $ be a `generalised' antisymmetric projection
operator which satisfies the relation
\beq
{\tilde P}_{i,i+1} K_{i, i+1} \, {\tilde \Lambda }  ~=~ - ~{\tilde \Lambda }~,
\label {b25}
\eeq
for all $i$. With the help of eqns.(\ref {b22}),
(\ref {b24}) and (\ref {b25}),  one can  prove that [32]
\beq
{\tilde \psi } ~=~
  \left [ \prod_{i<j} \sin { \pi \over L} (x_i -x_j)
\right ]^\beta \, {\tilde \Lambda } \,
\left (  \, \xi_{[\lambda ]} (z_1,z_2, \cdots ,z_N)  \,
\rho (\alpha_1 , \alpha_2 , \cdots , \alpha_N ) \, \right )
\label {b26}
\eeq
would be an eigenfunction of the spin CS Hamiltonian (\ref {b21})
with eigenvalue
$$\epsilon_{[\lambda ]} = { 2\pi^2 \over L^2 } \sum_{i=1}^N ~
\left [ \lambda'_i - \beta \left ( { N+1 \over 2 } -i \right ) \right ]^2
 .$$  Since this eigenvalue does not depend on the choice
of arbitrary function
$ \rho (\alpha_1 , \alpha_2 , \cdots , \alpha_N ) $, one usually
gets a set of degenerate eigenfunctions through the relation
(\ref {b26}).

Now, by combining (\ref {b18}) and (\ref {b25}), we see that the
antisymmetric projector
$ {\tilde \Lambda } $ satisfies the relation
$ \left ( Q_{i,i+1} P_{i,i+1} K_{i, i+1}  \right ) {\tilde \Lambda }  ~=~
- ~{\tilde \Lambda }~ $. So it is evident that the explicit form of
 $ {\tilde \Lambda } $  will depend on the choice of corresponding
 $Q_{i,i+1}$ matrix. For example, in the simplest case of a
spin CS model containing only two particles ($N=2$),
we  find that
 $ {\tilde \Lambda } = \one - K_{12} Q_{12} P_{12} $ satisfies the
relation (\ref {b25}).  By substituting this $ {\tilde \Lambda }$
to eqn.(\ref {b26}), we can write down the wave function for two particle
spin CS Hamiltonian as
\beq
{\tilde \psi } ~=~
  \left [  \sin { \pi \over L} (x_1 -x_2 )
\right ]^\beta \,  \left ( \one - K_{12} Q_{12} P_{12} \right ) \,
\left ( \, \xi_{[\lambda ]} (z_1,z_2) \,
\rho (\alpha_1 , \alpha_2 ) \, \right ) \, .
\label {b27}
\eeq
It is curious to notice that the $Q$ matrix,
originally appeared in the solution
(\ref {a5}) of YBE and the definition of extended $Y(gl_M)$ algebra,
also  plays an important role in constructing
the related wave function (\ref {b27}). So this
$Q$ matrix provides a direct link between the symmetry of
spin CS models and their exact wave functions.

It is worth noting that, while deriving the results of this section,
we have not used anywhere the power
series expansion (\ref {a7}) which is valid only for  the
$Q$ matrices associated with multi-parameter deformed $Y(gl_M)$
Yangian algebra. Therefore, our results would be equally applicable
for the case of multi-parameter deformed $Y(gl_M)$
Yangian symmetries  as well as  their nonstandard variants.
Thus, a spin CS Hamiltonian  like (\ref {b21})
would exhibit either of these two types of Yangian symmetries,
depending on the specific choice of corresponding $Q$ matrix.

\vspace{1cm}

\noindent \section { Conserved quantities and Lax
pairs of novel spin CS models }
\renewcommand{\theequation}{3.{\arabic{equation}}}
\setcounter{equation}{0}

\medskip

Though  in the previous section we have seen that
$T^0(u)$ matrix (\ref {b14}) generates the
conserved quantities of spin CS Hamiltonian
(\ref {b21}), we have not yet  derived those conserved quantities
in an explicit way. Our aim here is to write down
those conserved quantities in a compact form and find out
their connection with the
Lax pair of quantum integrable spin CS model (\ref {b21}).
For this purpose, we like to recall first the
procedure of constructing the conserved quantities of
usual spin CS model (\ref {a1}) from the related
Lax pair [8,3].  The Lax pair of  CS model (\ref {a1}) consists of
two $N\times N$ dimensional matrices
${\cal L}$ and  ${\cal M}$, whose  operator valued
elements are given by
\bea
&{\cal L}_{ij}  =  \, \delta_{ij} z_j { \d \over \d z_j } + \beta \,
\left ( 1 - \delta_{ij} \right ) \theta_{ij} P_{ij} \, , \, ~
{\cal M}_{ij}  = \,
 -  2 {\beta}'\, \delta_{ij} \sum_{k \neq i} h_{ik} P_{ik}
+ 2 {\beta'} \,
\left ( 1 - \delta_{ij} \right ) h_{ij} P_{ij} \, , ~~~~~~~ \nn \\
& \hskip 13.70 true cm  \nn
 (3.1a,b )
\eea
\addtocounter{equation}{1}
where ${\beta}' = {2\pi^2 \over L^2} \beta $,
$\theta_{ij} = { z_i \over z_i - z_j }$ and $h_{ij} =
\theta_{ij} \theta_{ji} $. It should be observed
that, unlike the case of previously discussed Lax operators
(\ref {b3}) or (\ref {b4}),
the above defined Lax pair does not depend on any auxiliary space
and can not give a solution of QYBE in a straightforward fashion.
However, through direct calculation
it can be checked
that, the Lax pair (3.1a,b) and the spin CS Hamiltonian
(\ref {a1}) obey the relations
\bea
&\left [ H  ,  {\cal L}_{ij} \right ] = \sum_{k=1}^N
\left (  {\cal L}_{ik} {\cal M}_{kj} -
 {\cal M}_{ik} {\cal L}_{kj}   \right ) \, , ~
\left [ H  ,  X_j^{\alpha \beta } \right ] = \sum_{k=1}^N
 X_k^{\alpha \beta } {\cal M}_{kj}  \, ,~ \sum_{k=1}^N
{\cal M}_{jk} = 0  \, ,  ~~~~~~~~~~~~~~~~~~~~~~~~~~~\nn \\
& \hskip 10.70 true cm
 \nn  (3.2a,b,c)
\eea
\addtocounter{equation}{1}
where $X_i^{\alpha \beta }$ denotes a spin dependent  operator
 which act as $ \v \alpha \r \l \beta \v $
on the spins of $i$-th particle and leave all other particles
untouched. By using eqn.(3.2a,b,c)
one can interestingly prove that the set of operators given by
\beq
T_n^{ \alpha \beta } ~=~ \sum_{i,j =1}^N  \,
 X_i^{\alpha \beta } \left ( {\cal L}^n \right )_{ij} \, ,
\label {c3}
\eeq
commute with the Hamiltonian (\ref {a1}).

The relation (\ref {c3}) gives an  explicit expression
 for the conserved quantities of spin CS Hamiltonian
(\ref {a1}) through the matrix elements of corresponding
${\cal L}$ operator.
However, from our discussions in the previous
section,  it is natural to expect that the $Q= \one $ limit
of $T^0(u)$ matrix (\ref {b14}) should also
produce these conserved quantities through some power
series expansion.  In fact it has been already shown that [3],
the $Q= \one $ limit of
projected monodromy matrix (\ref {b14})
can be expanded as
\beq
T^0(u) ~=~ \one \, + \, \beta \, \sum_{n=0}^\infty  \,
{1\over u^{n+1} } \, \sum_{\alpha , \beta = 1}^M  \,
\left ( X_0^{\alpha \beta } \otimes T_n^{\beta \alpha } \,
\right ) ,
\label {c4}
\eeq
 where $ T_n^{\alpha \beta }$s are given by
 (\ref {c3}). At present we like to construct
an analogue of eqn.(\ref {c4}) for a completely general
 $Q$ matrix.  Such a construction should yield
the conserved quantities of spin CS Hamiltonian
(\ref {b21}) and may also help to find out the related
Lax pair. In this context
it should be  noted that, for proving the
relation (\ref {c4}), a conjecture is made in ref.3
as
\beq
T^0(u) ~=~ \Pi \left (  \, \one + \beta \, \sum_{i=1}^N
{P_{0i} \over u- D_i } \, \right ) \, ,
\label {c5}
\eeq
where $\Pi ( K_{ij} ) = P_{ij} $,
$T^0(u)$ represents the $Q=\one $ limit of our monodromy matrix
(\ref {b14}) and $D_i$s are another type of Dunkl operators
given by
\beq
 D_i ~=~ z_i { \d \over \d z_i } ~+~
 \beta \, \sum_{j \ne i } \, \theta_{ij}  K_{ij}  ~ .
\label {c6}
\eeq
It is easy to check that these $D_i$s satisfy the relations
\beq
K_{ij} D_i ~= ~ D_j K_{ij} \, , ~~\left [ \, K_{ij} , D_k \,
 \right ] ~=~ 0 \, , ~~\left [ \, D_i , D_j \, \right ] ~=~
\beta \, \left ( D_i -D_j \right ) K_{ij} \, ,
\label {c7}
\eeq
where $k\neq i,j $. We propose now a generalisation of
the conjecture (\ref {c5}) as
\beq
T^0(u) ~=~ \Pi^* \left (  \, \Omega + \beta \sum_{i=1}^N
{\P_{0i} \over u- D_i } \, \right ) \, ,
\label {c8}
\eeq
where $T^0(u)$ is defined by (\ref {b14}),  and
\bea
~\Omega ~ = ~ Q_{01} Q_{02} \cdots Q_{0N} \, , ~~
\P_{0i} ~ = ~ \left ( Q_{01} Q_{02} \cdots Q_{0,i-1} \right ) \, P_{0i} \,
\left ( Q_{0,i+1} Q_{0,i+2} \cdots Q_{0N} \right )   \, . \nn
~~~(3.9a,b)
\eea
\addtocounter{equation}{1}
By using the relations (\ref {a6}), (\ref {b18}) and (\ref {b19}),
we have checked the validity of above
conjecture for systems containing small number of particles.
Moreover, it is evident that at the limit $Q= \one $ (when
one can  put $\Pi^* = \Pi , ~
\Omega ~=~\one ,$ and $ \P_{0i} = P_{0i} $), equation
 (\ref {c8}) reproduces the previous conjecture (\ref {c5}).
So, in the following,  we shall assume that
(\ref {c8}) is a valid relation for
any possible choice of corresponding
$Q$ matrix and all values of particle number $N$.

Next, we like to derive two relations which will be used shortly
 to express the conjecture (\ref {c8}) in a  more convenient form.
First of all, by applying eqns.(\ref {c6}), (\ref {c7}) and (\ref {b9}),
 one can show that
\beq
\Pi^* \left ( D_i^n \right ) ~=~ \sum_{j=1}^N \,
\left (  {\tilde {\cal L}}^n  \right )_{ij} \, ,
\label {c10}
\eeq
where $ {\tilde {\cal L}}$ is a $N \times N$ matrix with elements
given by
\beq
{\tilde {\cal L}}_{ij}
~=~ \delta_{ij} z_j { \d \over \d z_j } ~+~ \beta \,
\left ( 1 - \delta_{ij} \right ) \theta_{ij} {\tilde P_{ij} } \, .
\label {c11}
\eeq
It may be noted that (\ref {c10}) is  a
straightforward generalisation of the known relation [3]:
$\Pi \left ( D_i^n \right ) ~=~ \sum_{j=1}^N \,
\left ( {\cal L}^n  \right )_{ij} $.
Secondly, by using the standard
relation: $ P_{0i} ~=~ \sum_{\alpha , \beta = 1}^M \,
X_0^{\alpha \beta } \otimes  X_i^{\beta \alpha } \,$
and the conditions  (\ref {a6}),
we find that the operator
$\P_{0i}$ (3.9b)  can be rewritten as
\beq
\P_{0i} ~=~ \sum_{\alpha , \beta = 1}^M \,
X_0^{\alpha \beta } \otimes {\tilde X}_i^{\beta \alpha } \, ,
\label {c12}
\eeq
where
\beq
 {\tilde X}_i^{\beta \alpha } ~=~ Q_{i,i+1} Q_{i,i+2} \cdots
 Q_{iN}  \, X_i^{\beta \alpha } \,
  Q_{i1} Q_{i2} \cdots Q_{i,i-1} \, .
\label {c13}
\eeq
The expression (\ref {c12}) is more suitable
for our purpose than (3.9b),
since in (\ref {c12}) we have only one operator
$X_0^{\alpha \beta }$  which depends on the $0$-th auxiliary space.

Now,
with the help of eqns.(\ref {c10}) and (\ref {c12}),
we find that the relation (\ref {c8})
can be expressed in a nice form 
\beq
T^0(u) ~=~ \Omega \, + \, \beta \, \sum_{n=0}^\infty  \,
{1\over u^{n+1} } \, \sum_{\alpha ,\beta = 1}^M  \,
\left (\, X_0^{\alpha \beta } \otimes
{\tilde T}_n^{\beta \alpha } \, \right )  ~ ,
\label {c14}
\eeq
 where $ {\tilde T}_n^{\alpha \beta } $s are given by
\beq
{\tilde T}_n^{ \alpha \beta } ~=~ \sum_{i,j=1}^N \,
 {\tilde X}_i^{\alpha \beta }
\left ( {\tilde {\cal L}}^n \right )_{ij} \, .
\label {c15}
\eeq
It is worth noting that eqns.(\ref {c14}) and (\ref {c15})
give us the desired generalisation of previous relations (\ref {c4})
and (\ref {c3}), for the case of an arbitrary $Q$ matrix.
So the operators  $ {\tilde T}_n^{ \alpha \beta }$ represent the
conserved quantities of spin
CS Hamiltonian (\ref {b21}), for any possible choice of
corresponding $Q$ matrix. It may also be observed that
the operator $ \Omega $, appearing in eqn.(\ref {c14}), would generate
$M^2$ number of additional conserved quantities (all of
 which become trivial at $Q= \one $ limit).
However, due to the fact
that $ \Omega $ (3.9a) depends on the $0$-th auxiliary space
in a very complicated way, we find it difficult to write
down the explicit form of these
$M^2$ number of additional conserved quantities.

It may be noticed that
the above discussion, which yields the explicit form of conserved
quantities  (\ref {c15}),
heavily depends on our conjecture (\ref {c8}).
So, in the following,  we like to show in an independent way that
$ {\tilde T}_n^{ \alpha \beta }$s are indeed  conserved
quantities for the spin CS Hamiltonian (\ref {b21}).
For this purpose we first compare the two expressions
(\ref {c3}) and (\ref {c15}). Such
comparison clearly indicates that
the operator ${\tilde {\cal L}}$ (\ref {c11}) may  be treated as
a generalisation of ${\cal L}$ (3.1a), for the case of an
arbitrary $Q$ matrix. Moreover, it is evident that one can produce
the matrix elements of ${\tilde {\cal L}}$ from that of ${\cal L}$
(3.1a),  through the simple substitution:
$ P_{ij} \rightarrow {\tilde P}_{ij}$.
So we make a similar substitution to eqn.(3.1b)
and  write down the matrix elements of corresponding
${\tilde {\cal M}}$ as
\beq
{\tilde {\cal M}}_{ij} ~=~ - \,  2 \beta' \, \delta_{ij}
\sum_{k \neq i} \left ( h_{ik} {\tilde P}_{ik} \right )
~+~ 2 \beta' \,
\left ( 1 - \delta_{ij} \right ) \, h_{ij} {\tilde P}_{ij} \, .
\label {c16}
\eeq
Now, we interestingly  find that the four operators
${\tilde H}$, ${\tilde {\cal L}}$,  ${\tilde {\cal M}}$ and
 ${\tilde X}_i^{\beta \alpha }$,  given by
equations (\ref {b21}), (\ref {c11}),  (\ref {c16})
and (\ref {c13}) respectively,
satisfy the relations
\bea
&~~~~~~~~~~~~~
\left [ \, {\tilde H} \, , \, {\tilde {\cal L}}_{ij} \, \right ] ~=~
\sum_{k=1}^N \, \left ( \, {\tilde {\cal L}}_{ik}
{\tilde {\cal M}}_{kj} \, - \,
 {\tilde {\cal M}}_{ik} {\tilde {\cal L}}_{kj}  \, \right ) \, , ~~
\nn ~~~~~~~~~~~~~~~~~~&~~(3.17a) \\
&~~~~~~~~~~~~~
\left [ \, {\tilde H} \, , \, {\tilde X}_j^{\alpha \beta } \, \right ]
~=~ \sum_{k=1}^N \,
 {\tilde X}_k^{\alpha \beta } {\tilde {\cal M}}_{kj}
\, , ~~~~ \sum_{k=1}^N \,
{\tilde {\cal M}}_{jk} ~=~ 0  \, . ~\nn
~~~~~~~~~~~~~~~~~&(3.17b,c)
\eea
\addtocounter{equation}{1}
Again, these relations are a straightforward generalisation of the
previous equation (3.2a,b,c). In fact the eqns.(3.2a,b,c) and
(3.17a,b,c) are exactly same in form and  related to each other
through the substitutions $ {\cal L} \leftrightarrow
 {\tilde {\cal L}}$, $ {\cal M} \leftrightarrow  {\tilde {\cal M}}$,
 $ X_k^{\alpha \beta }  \leftrightarrow {\tilde X}_k^{\alpha \beta }$
and $H \leftrightarrow {\tilde H}$. So it is clear that
${\tilde {\cal L}}$ (\ref {c11})
and  ${\tilde {\cal M}}$ (\ref {c16}) represents the
Lax pair associated with the spin CS Hamiltonian ${\tilde H}$
(\ref {b21}).  Furthermore,
by using relations (3.17a,b,c),  it is easy
to directly check that the operators
 ${\tilde T}_n^{\alpha \beta } $ (\ref {c15}) commute with
${\tilde H}$. Thus we are able to prove in an independent
way that
 $ {\tilde T}_n^{\alpha \beta }$s are the conserved quantities
of  spin CS Hamiltonian (\ref {b21}) and  find out
how these conserved quantities are related to the
corresponding Lax pair.

\vspace{1cm}

\noindent \section {Specific examples of spin CS models
with extended $Y(gl_M)$ Yangian symmetry}
\renewcommand{\theequation}{4.{\arabic{equation}}}
\setcounter{equation}{0}

\medskip

In the previous sections we have developed a rather
general framework for constructing the spin CS Hamiltonian which
would exhibit an extended $Y(gl_M)$
Yangian symmetry, and also found out the
related Lax pair as well as conserved quantities.
Let us  derive now a few
particular solutions of YBE  which can be expressed
in the form (\ref {a5}) and subsequently apply
our general results to obtain the concrete form of
corresponding spin CS models, Lax pairs etc.

\smallskip

\noindent {\it Case 1.}

To generate a rational $R$ matrix of the form (\ref {a5}),
one may use the well known spectral parameter independent solution
of YBE (\ref {a4}) given by
\beq
R_{00'} ~=~ \sum_{\sigma = 1}^M ~   \epsilon_\sigma (q) ~
e^0_{\sigma \sigma }  \otimes   e^{0'}_{\sigma \sigma } ~+~
\sum_{ \sigma \neq \gamma } ~e^{ i \phi_{ \gamma  \sigma } }~
e^0_{\sigma \sigma } \otimes  e^{0'}_{\gamma \gamma } ~+~
\left ( q - q^{-1} \right )~ \sum_{  \sigma  < \gamma } ~
e^0_{\sigma \gamma } \otimes  e^{0'}_{\gamma \sigma } ~,
\label {d1}
\eeq
where $ e^0_{\sigma \gamma } $ is a  basis operator on the $0$-th
auxiliary space with elements
$ \left ( e^0_{\sigma \gamma } \right )_{ \tau \delta }
= \delta_{\sigma \tau } \delta_{ \gamma \delta } $,
$~\phi_{ \gamma  \sigma }$s are ${ M(M-1) \over 2 }$ number of
independent antisymmetric deformation parameters:
$ \phi_{ \gamma  \sigma }  =  - \phi_{ \sigma  \gamma } $,
and each of the $  \epsilon_\sigma (q) $ can be   freely
taken as either $ q$ or $- q^{-1} $ for any value of $\sigma $.
In the special  case  when
all  $\epsilon_\sigma (q)$s  take the same value
(i.e., all of them are either $q$ or $-q^{-1}$),  (\ref {d1})
can be obtained from the universal
${\cal R}$-matrix associated with  $U_q(sl(M))$ quantum group, for generic
values of the parameter $q$ [18-19].
On the other hand if
$\epsilon_\sigma (q)$s  do not take the same value for all $\sigma $,
the corresponding `nonstandard' solutions  are found to be connected with
the universal ${\cal R}$ matrix of $U_q(sl(M))$ quantum group,
when $q$ is a root of unity [33-35]. It may also be noted that
 the parameters $ \phi_{ \sigma \gamma }$
and $\epsilon_\sigma (q)$  have appeared
previously in the context of multi-parameter dependent quantisation
of $GL(M)$ group [36] and some asymmetric vertex models [37].
It is now easy to check directly
  that the $R_{00'}$ matrix (\ref {d1}) satisfies the  condition:
$R_{00'} -  P_{00'} (R_{00'})^{-1}  P_{00'} = (q-q^{-1}) P_{00'}$.
So, by  following the Yang-Baxterisation prescription [38]
related to Hecke algebra, we can  construct a
 spectral parameter dependent solution of YBE (\ref {a4}) as
\beq
 R_{00'}(u)~=~q^{u \over \beta } \,  R_{00'} \, - \,
q^{-{ u \over \beta }} \,  P_{00'} (R_{00'})^{-1}  P_{00'} \, .
\label {d2}
\eeq
Substituting
the explicit form of $ R_{00'}$ (\ref {d1}) to  the above expression,
multiplying it by the constant
 $ \beta / ( q-q^{-1}) $ and subsequently taking  the
$q \rightarrow 1$ limit,
we get a rational solution of YBE in the
form (\ref {a5}) where the corresponding $Q$ matrix is given by
\beq
Q_{00'}~=~ \sum_{\sigma =1}^M \, \epsilon_{\sigma }
 \, e^0_{ \sigma \sigma }\otimes e^{0'}_{ \sigma \sigma }~+~
\sum _{ \sigma \neq \gamma } \, e^{i\phi_{\gamma \sigma }}
 \, e^0_{ \sigma \sigma }\otimes e^{0'}_{ \gamma \gamma } \, ,
\label {d3}
\eeq
$\epsilon_{\sigma } $s being $M$ number of discrete parameters,
each of which can be freely chosen as $1$ or $-1$.
One may also verify directly  that the above $Q$ matrix
satisfies the two required  conditions (\ref {a6}).
Consequently, the rational solution of YBE associated with
this $Q$ matrix can be used to
define a class of extended $Y(gl_M)$ Yangian algebra. Moreover,
it is  worth
observing that only for the special choice
$\epsilon_{\sigma } = 1$ (or, $\epsilon_{\sigma } = - 1$)
 for all $\sigma $,
the $Q$ matrix (\ref {d3}) admits an expansion in the form
(\ref {a7}). Therefore, only for these two choices of discrete
parameters, the corresponding $Q$ matrix  generates
a multi-parameter dependent deformation of $Y(gl_M)$ Yangian
algebra [21,22]. For all other choices of discrete parameters
$\epsilon_{\sigma } $, we would get some nonstandard variants
of $Y(gl_M)$ Yangian algebra.

Next we substitute the specific form of $Q$ matrix
(\ref {d3}) to eqn.(\ref {b18}) and find that
\beq
{\tilde P}_{i,i+1}~=~ \sum_{\sigma =1}^M \, \epsilon_{\sigma }
 \, e^i_{ \sigma \sigma }\otimes e^{i+1}_{ \sigma \sigma }~+~
\sum _{ \sigma \neq \gamma } \, e^{i\phi_{\gamma \sigma }}
 \, e^i_{ \sigma \gamma }\otimes e^{i+1}_{ \gamma \sigma  } \, .
\label {d4}
\eeq
>From our discussion in sec.2 it is evident that,
any particular choice of $\epsilon_{\sigma } $s and
$\phi_{\gamma \sigma }$s in the above expression of
$ {\tilde P}_{i,i+1}$  would give us a representation of
the permutation algebra (2.11). The action
of ${\tilde P}_{i,i+1} $ (\ref {d4}) on the space ${\cal F}$
can easily be written as
\beq
{\tilde P}_{i,i+1} \,
 \v \, \alpha_1 \alpha_2 \cdots \alpha_i \alpha_{i+1}
\cdots \alpha_N  \, \r ~~=~~
 \exp \left ( i\phi_{\alpha_i \alpha_{i+1} } \right ) \,
\v \, \alpha_1
\alpha_2 \cdots \alpha_{i+1} \alpha_i  \cdots  \alpha_N \, \r   , ~~
\label {d5}
\eeq
where we have  used the notation
$e^{i \phi_{\sigma  \sigma }} = \epsilon_\sigma $. It is
interesting to observe that, the above
`anyon like' representation of permutation
algebra not only interchanges the spins of two particles
but also picks up an appropriate phase factor.
Moreover, by substituting (\ref {d4}) to
(\ref {b19}) one can find out the  operators
${\tilde P}_{ij}$, when $j-i >1$. The action of such
an operator on the space ${\cal F}$ is given by
\bea
&&{\tilde P}_{ij} \,
 \v \alpha_1 \alpha_2 \cdots \alpha_i \cdots \alpha_j
\cdots \alpha_N \r ~~=~~~~~~~~~~~~~~~~~~~~~~~~~~~~~ \nn \\ &&
 \quad \quad \quad
 \exp   \left \{  \,  i \, \phi_{\alpha_i \alpha_j} ~+~ i \,
\sum_{\tau =1}^M  \, n_\tau \,   \left (   \phi_{\tau \alpha_j} -
 \phi_{\tau \alpha_i}    \right )  \,  \right \}   \,
     \v \alpha_1
\alpha_2 \cdots \alpha_j \cdots \alpha_i \cdots \alpha_N \r   , ~~
\label {d6}
\eea
where $n_\tau $ denotes the number of times of occuring
 the  particular spin orientation $ \tau $
in the configuration
$ \v \alpha_1  \cdots \alpha_i \cdots
\alpha_p \cdots \alpha_j \cdots  \alpha_N \r $,
when the index $p$ in $\alpha_p $ is varied from $i+1$ to $j-1$.
Thus, it turns out that the phase factor  associated with the
element ${\tilde P}_{ij}$ actually depends on the spin configuration
of $(j-i+1)$ number of particles.
Consequently
the operator ${\tilde P}_{ij}$, which acts nontrivially on the spin space
of all these $(j-i+1)$ number of particles, would generate  a
highly nonlocal  many-body  spin dependent interaction
in the  CS Hamiltonian (\ref {b21}).
Evidently at the special case $\epsilon_\sigma = 1 $ and
$ \phi_{ \gamma  \sigma }  = 0$ for all $\sigma ,\,  \gamma $,
this ${\tilde P}_{ij}$  reduces to two-body spin dependent interaction
$P_{ij}$ (\ref {b13}), which is used to define the usual spin CS
model (\ref {a1}).

It may be noted that the `anyon like' representations
((\ref {d5}),(\ref {d6})) and the related spin CS Hamiltonians
were considered earlier in ref.32. However, through our
present analysis, we are able to construct these
anyon like representations in a systematic way from the given
solutions of YBE associated with $Q$ matrix (\ref {d3}).
Moreover we are able to show that
the spin CS Hamiltonian (\ref {b21}), which contains these
${\tilde P}_{ij}$ ((\ref {d5}),(\ref {d6}))
as spin dependent interaction,
would exhibit the extended $Y(gl_M)$ Yangian symmetry generated
through $Q$ matrix (\ref {d3}). Furthermore,  by substituting
(\ref {d3}), (\ref {d5}) and (\ref {d6}) to
eqns.(\ref {c11}), (\ref {c16}) and (\ref {c15}), one can
explicitly find out the corresponding Lax pair and conserved
quantities.

It is important to note that,
we can change the symmetry algebra of
spin CS Hamiltonian (\ref {b21}) by tuning the discrete parameters
$ \epsilon_\sigma $ and continuous parameters
$ \phi_{\gamma \sigma } $ in the related $Q$ matrix (\ref {d3}).
Therefore, the study of corresponding degenerate wave functions
 should give us valuable information about
the representation theory of a large class of extended Yangian
algebras. In the following we like to construct the ground states
of above considered spin CS models, for the simplest case
when they contain  only two spin-${1\over 2}$ particles ($N=M=2$),
and examine the dependence of these ground states
on the related discrete as well as continuous deformation parameters.
Indeed, by using eqn.(\ref {b23}),   it  is easy to see
that the trivial monomial $\xi (z_1 , z_2 ) =1 $
would be an eigenvector of two Dunkl operators
$d_1, \, d_2 $ and will also correspond to
the lowest eigenvalue ($ \beta^2/2 $) of operator
${\cal H}^* = d_1^2 + d_2^2 $.
Therefore, by  substituting $\xi (z_1 , z_2 ) =1 $ to
eqn.(\ref {b27}), the ground state
associated with energy eigenvalue ${\pi^2 \beta^2 \over  L^2 } $
can be obtained as
\beq
{\tilde \psi} ~~=~~\sin^\beta
 \left \{  { \pi \over L } (x_1 -x_2)   \right \}  ~
\left ( 1 - Q_{12} P_{12}  \right ) \rho( \alpha_1 , \alpha_2 ) ~.
\label {d7}
\eeq
Moreover, for spin-${1\over 2}$  case,
the arbitrary spin dependent function $\rho (\alpha_1 , \alpha_2 ) $
can be chosen in four different ways: $\v 11 \r $,
$ \v  12 \r$,  $\v  21 \r$  and  $ \v 22 \r $. Inserting these
forms of $\rho$ to eqn.(\ref {d7}) and also using eqn.(\ref {d5})
for  $N=M=2$ case,
 we get three degenerate eigenfunctions like
$$
{\tilde \psi}_1  \, = \, ( 1 - \epsilon_1 ) \, \Gamma^\beta \v 11 \r~,~~
{\tilde \psi}_2 \, = \, ( 1 - \epsilon_2 ) \, \Gamma^\beta \v 22 \r~,~~
{\tilde \psi}_3 \, = \, \Gamma^\beta ~\left ( \v 12 \r - e^{i \theta }
 \v 21 \r \right )~,~ \eqno  (4.8a,b,c)
$$
\addtocounter{equation}{1} \noindent
where $ \Gamma = \sin \left \{
  { \pi \over L } (x_1 -x_2) \right \} $ and $\theta = \phi_{12} $.
Notice that the choice of
$\rho $ as  $ \v 12 \r $ or  $\v 21 \r$ would  lead to
 the same wave function  $ {\tilde \psi}_3  $ up to a multiplicative constant.
Now it may be observed that, the substitution
 $ \epsilon_1 = \epsilon_2 = 1$ and $\theta = 0 $ to eqn.(4.8)
would give us the ground state wave function of usual $Y(gl_2)$ symmetric
 spin CS model (\ref {a1}) when $N=M=2$. However,  only the
 wave function ${\tilde \psi}_3$ in eqn.(4.8) remains nontrivial
after the above mentioned substitution, which
 in this case  actually gives a nondegenerate ground state.
In a similar way one finds
 a nondegenerate ground state even for the slightly different case:
 $ \epsilon_1 = \epsilon_2 = 1$ and $\theta \neq 0 $.
But, it should be noticed that the choice
 $ \epsilon_1 = \epsilon_2 = 1$ and $\theta \neq 0 $
leads to a spin CS Hamiltonian of type (\ref {b21})
whose symmetry algebra is given by  a one parameter deformation of
 $Y(gl_2)$ Yangian. Therefore, it is apparent that the change of usual
 $Y(gl_2)$ Yangian symmetry of a spin CS model,
 through a continuous deformation parameter $\theta $,
does not affect the degeneracy factor of related ground
states. On the other hand if one substitutes
$ \epsilon_1 = - \epsilon_2 = 1$ to eqn.(4.8), then both
$ {\tilde \psi}_2 $ and $ {\tilde \psi}_3 $
would remain nontrivial
and, as a result, we will get a doubly degenerate ground state.
But, from our previous discussion it is known that the choice
$ \epsilon_1 = - \epsilon_2 = 1$ in $Q$ matrix (\ref {d3})
leads to a nonstandard variant of $Y(gl_2)$ Yangian.
So we curiously find that, by switching over to
a nonstandard variant of $Y(gl_2)$ Yangian symmetry from its
standard counterpart, one can change the degeneracy factor
of the related ground states.

\smallskip

\noindent {\it Case 2. }

Recently, some new rational solutions of YBE
is constructed  from the universal ${\cal R}$
matrices of deformed Yangian algebras [24].
In particular, an explicit solution
is found in the form (\ref {a5}) where the corresponding $Q$ matrix
is given by
\beq
    Q_{00'} ~=~ \one \, + \, 2 \xi \,
    r_{00'}  \, + \, 2 \xi^2  \, (r_{00'})^2 \, ,
\label {d9}
\eeq
with
\bea
    &r_{00'}  ~=~ { 1\over 2} \sum_{\sigma < M+1- \sigma }
\left ( h^0_\sigma \otimes e^{0'}_{ \sigma , M+1 - \sigma }
- e^{0}_{ \sigma , M+1 - \sigma } \otimes h^{0'}_\sigma \right )~~~~~ \nn \\
& \hskip 5 true cm   + \sum_{ \sigma < \gamma < M+1 - \sigma  }
\left ( e^0_{\sigma \gamma } \otimes e^{0'}_{ \gamma , M+1 - \sigma }
- e^{0}_{ \gamma , M+1 - \sigma } \otimes e^{0'}_{ \sigma \gamma }
 \right ) \, , \nn
\eea
\noindent
and $ h^0_\sigma =
 e^0_{\sigma \sigma } -  e^{0}_{ M+1 - \sigma , M+1 - \sigma } $.
Since this Q matrix admits an expansion like
(\ref {a7}),  the corresponding rational
solution of YBE would generate a single parameter dependent deformation of
$Y(gl_M)$ Yangian algebra. Again, by substituting the $Q$ matrix
(\ref {d9}) to eqns.(\ref {b18}) and (\ref {b19}), one can
 construct a new representation of permutation algebra (\ref {b10}).
Evidently, this representation of permutation algebra can be used
to find out a spin CS Hamiltonian like (\ref {b21}),
which would exhibit the above mentioned
single parameter deformed $Y(gl_M)$ Yangian symmetry. However,
it is rather difficult to explicitly write down
such representation of permutation algebra due to its complicated nature,
and we present here only the action of ${\tilde P}_{12}$
for $N=M=2$ case:
\bea
&{\tilde P}_{12} \, \v  11 \r ~=~ \v 11 \r \, ,
~~~~ {\tilde P}_{12} \, \v 12 \r ~=~ \v  2 1  \r \, - \, \xi  \,
\v   1 1  \r \, , \hskip 4 true cm & \nn \\
&{\tilde P}_{12} \,  \v 21 \r ~= ~ \v 12 \r  +  \, \xi \, \v 11 \r  \, , ~~~
{\tilde P}_{12} \, \v 22 \r ~=~  \v 22 \r
+ \,  \xi \, \v 12 \r   - \,  \xi \, \v 21 \r
+ \,  \xi^2 \, \v  1 1 \r \,  .~~~~ &
\label {d10}
\eea
Remarkably, this
${\tilde P}_{12}$ can create new spin components
which are not present in the original spin configuration.
By substituting  the $Q$ matrix (\ref {d9})
and corresponding ${\tilde P}_{ij}$  operators to
eqns.(\ref {c11}), (\ref {c16}) and (\ref {c15}),
in principle one can also
find out the related Lax pair as well as conserved
quantities. Moreover,
by  using eqns.(\ref {d10}) and (\ref {d7}), it is easy to
construct the ground state of associated
spin CS model (\ref {b21}) for $N=M=2$ case as
\beq
{\tilde \psi} ~~=~~\sin^\beta
 \left \{  { \pi \over L } (x_1 -x_2)   \right \}  \,
\left(\, \v 12 \r - \v 21 \r + 2 \xi  \v 11 \r  \,\right) \, .
\label {d11}
\eeq
Thus we get here a nondegenerate ground state, which reduces
 to the ground state of usual $Y(gl_2)$ symmetric spin CS
model at $\xi = 0$ limit. Thus we see again that,  change
of Yangian symmetry through a continuous deformation parameter
does not affect the degeneracy  of related ground state.

\smallskip
\noindent {\it Case 3.}

We propose another rational solution of YBE which can be
expressed in the form (\ref {a5}), where the corresponding
$Q$ matrix is given by
\beq
  Q_{00'} ~=~ \one + \xi   \sum_{\sigma =2 }^M
\left ( e^0_{\sigma \sigma } \otimes e^{0'}_{ 1  \sigma }
- e^{0}_{ 1  \sigma } \otimes e^{0'}_{\sigma \sigma } \right )
 \, .
\label {d12}
\eeq
 It is clear that this $Q$ matrix would generate a new type of
single parameter deformed $Y(gl_M)$ Yangian algebra. Moreover,
by substituting this $Q$ matrix to
 eqn.(\ref {b18}), it is possible to construct
a representation of permutation algebra (2.11) as
\beq
  {\tilde P}_{i,i+1} ~=~
   P_{i,i+1}  + \xi \sum_{\sigma =2 }^M
\left ( e^i_{\sigma \sigma } \otimes e^{i+1}_{ 1  \sigma }
- e^{i}_{ 1  \sigma } \otimes e^{i+1}_{\sigma \sigma } \right )
 \, ,
\label {d13}
\eeq
where $e^i_{ \sigma \gamma } \equiv X_i^{\sigma \gamma }$.
For the simplest $N=M=2$ case, the action of above permutation operator
may be written as
\beq
{\tilde P}_{12} \v 11 \r = \v 11 \r \, , ~
{\tilde P}_{12} \v 12 \r = \v 21 \r \, , ~
{\tilde P}_{12} \v 21 \r = \v 12 \r \, , ~ {\tilde P}_{12} \v 22 \r =
\v  22  \r  - \xi  \v 12 \r + \xi  \v 21 \r \, .
\label {d14}
\eeq
By substituting (\ref {d13}) to (\ref {b19}),  one can also
find out the operators ${\tilde P}_{ij}$,
when $j-i >1$. Evidently, these permutation operators will give us
a spin CS Hamiltonian of the form
(\ref {b21}),  which would exhibit a deformed $Y(gl_M)$
Yangian symmetry related to the $Q$ matrix (\ref {d12}).
Again, in principle, we  can  explicitly construct the  Lax pair
as well as conserved quantities for such spin CS Hamiltonian,
 by inserting the $Q$ matrix (\ref {d12}) and associated
${\tilde P}_{ij}$ operators to the general relations
(\ref {c11}), (\ref {c16}) and (\ref {c15}). Moreover,
with the help of eqns.(\ref {d7}) and
(\ref {d14}),  it is rather easy to see that the corresponding
nondegenerate  ground state (for $N=M=2$ case) would actually
coincide with the ground state of usual $Y(gl_2)$ symmetric
spin CS model.

\smallskip

\noindent {\it Case 4.}

Finally we indicate about a particular class of possible $Q$ matrix
solutions, which using (\ref {a5}) yields rational solutions of YBE
and through (\ref {b18}) constructs novel spin CS models (\ref {b21}).
Such $Q$ matrix solutions may be given as
\beq
  {Q}_{i,i+1} ~=~
   F_{i,i+1} F_{i+1,i}^{-1}
 \, ,
\label {d15}
\eeq
  where $ F_{i,i+1}$ are some representations of 
twisting operators defined in ref.39.
Remarkably, the necessary conditions for the twisting operators can be 
shown also to be  
sufficient for the  $Q$-matrix  (\ref{d15}) as a solution of
(\ref{a6}) [24]. 
Few concrete examples of such twisting matrices  may be given as
$$ \mbox{i)}~ F_{i, i+1}=
 \one + \xi  \sigma_3^i
 \otimes  \sigma^{i+1}_-
~~ ~\mbox {and ~~~ii)~}  F_{i, i+1}=
\one + \xi   \sum_{\alpha =2 }^M
 e_{\alpha\alpha }^{i} \otimes e^{i+1}_{ 1  \alpha } \, .
$$

For some twisting operators  the construction  as
$$  {Q}_{i,i+1} ~=~
   F_{i,i+1} \Omega_{i,i+1}F_{i+1,i}^{-1} $$ with 
$
\Omega_{i,i+1}=~ \sum_{\sigma =1}^M \, \epsilon_{\sigma }
 \, e^{i}_{ \sigma \sigma }\otimes e^{i+1}_{ \sigma \sigma }~+~
\sum _{ \sigma \neq \gamma } \, 
 \, e^i_{ \sigma \sigma }\otimes e^{i+1}_{ \gamma \gamma } \, $
may generate  $Q$-matrices 
corresponding to the nonstandard Yangian algebras.  
Such an example of the twisting operator is 
$$F_{ii+1}  ~=~\exp\left [ { i} \sum_{\sigma  \neq \gamma }
 h^i_\sigma \otimes h^{i+1}_\gamma \phi_{\sigma \gamma}
\right ],~~$$
where $ h^i_\sigma =
 e^i_{\sigma \sigma } -  e^{i}_{ \sigma +1,  \sigma +1 } $ and $
\phi_{\sigma \gamma}
$ are deforming parameters with $
\phi_{ \gamma \sigma}
=-\phi_{\sigma \gamma} .$

\vspace{1cm}

\noindent \section { Concluding Remarks }

\medskip

In this article we have constructed the general form of
spin Calogero-Sutherland (CS) model which would satisfy
the extended (i.e., multi-parameter dependent including
 nonstandard variant of)
$Y(gl_M)$ Yangian symmetry.
An important feature of such CS models is that they
contain spin dependent   many-body type interactions , which can be calculated
directly from the associated rational solutions of Yang-Baxter equation.
More interestingly these spin dependent interactions can  be expressed
through some novel representations of permutation algebra on the
combined internal space of all
particles. We have also
established the integrability  by  finding
 out the general forms of conserved quantities
and Lax pairs for this class of spin CS models.
As fruitful applications of the formalism  we have 
constructed some concrete examples of
spin CS models which exhibit the extended $Y(gl_M)$ Yangian symmetry
and discussed about the structure of the 
related ground state wave functions. Finally we have indicated the
possible connections
with  twisting operators  in some particular cases.

The existence of extended $Y(gl_M)$
Yangian symmetry in the above mentioned class of spin CS models
might lead to interesting
applications in several directions. As it is well
known, the degeneracy of wave functions for usual spin
CS model can be explained quite nicely through the representations of
$Y(gl_M)$ algebra.  So it is natural to expect that the representations
of extended $Y(gl_M)$ algebra would play a similar role
in identifying the degenerate multiplates of corresponding spin CS models.
Conversely, one may also be able to extract
valuable information about the representations of extended
$Y(gl_M)$ Yangians, by studying the wave functions of associated
spin CS models. In this article we have analysed the degeneracy of a few
ground state wave functions, which indicated that the representations
of nonstandard variants of $Y(gl_M)$ Yangian algebra may differ considerably
from their standard counterpart.
In particular it has been found that the continuous deformation by 
multiparameters seems not to change the degeneracy pattern. However the
nonstandard cases with discrete change of symmetries affect the degeneracy
picture with a tendency of creating more degenerate states.
 
The represenatations
of nonstandard variants of $Y(gl_M)$ Yangian algebra might turn out to
be a rather interesting subject for future investigation.
Moreover, one may also try to use the extended $Y(gl_M)$ symmetry
in spin CS models for calculating their
dynamical correlation functions and various thermodynamic
properties. Finally, we hope that it would be possible
to find out many other new type of quantum
integrable systems with long range interactions,
which would exhibit the extended $Y(gl_M)$ Yangian symmetry.

\newpage
\leftline {\large \bf References }
\medskip
\begin{enumerate}

\item  M.A. Olshanetsky and A.M. Perelomov, Phys. Rep. 94 (1983) 313.

\item  F.D.M. Haldane, Z.N.C. Ha, J.C. Talstra, D. Bernerd and V. Pasquier,
       Phys. Rev. Lett. 69 (1992) 2021.

\item  D. Bernard, M. Gaudin, F.D.M. Haldane and V. Pasquier,
       J. Phys. A 26 (1993) 5219.

\item  A.P. Polychronakos, Phys. Rev. Lett. 69 (1992) 703.

\item  B. Sutherland and B.S. Shastry, Phys. Rev. Lett. 71 (1993) 5.

\item  M. Fowler and J.A. Minahan, Phys. Rev. Lett. 70 (1993) 2325.

\item  L. Brink, T.H. Hansson and M. Vasiliev, Phys. Lett. B 286 (1992) 109.

\item  K. Hikami and M. Wadati, J. Phys. Soc. Jpn. 62 (1993) 4203;
       Phys. Rev. Lett. 73 (1994) 1191.

\item  H. Ujino and M. Wadati, J. Phys. Soc. Jpn. 63 (1994) 3585;
       J. Phys. Soc. Jpn. 64 (1995) 39.

\item  K. Hikami, Nucl. Phys. B 441 [FS] (1995) 530.

\item  F.D.M. Haldane, in Proc. 16th Taniguchi Symp., Kashikijima,
       Japan, (1993) eds. A. Okiji and N. Kawakami (Springer, Berlin,
       1994).

\item  A. Cappelli, C.A. Trugenberger, and G.R. Zemba, Phys. Rev. Lett.
       72 (1994) 1902.

\item  M. Stone and M. Fisher, Int. J. Mod. Phys. B 8 (1994) 2539.

\item  F. Lesage, V. Pasquier and D. Serban, Nucl. Phys. B 435 [FS] (1995)
       585.

\item  H. Awata, Y. Matsuo, S. Odake and J. Shiraishi, Nucl. Phys. B 449
       (1995) 347.

\item  J. Avan, A. Jevicki, Nucl. Phys. B 469 (1996) 287.

\item  J. Avan, A. Jevicki and J. Lee, {\it Yangian-invariant
       field theory of matrix-vector models }, PAR LPTHE 96-22,
       hep-th/9607083.

\item  V.G. Drinfeld, {\it Quantum Groups }, in ICM proc. (Berkeley, 1987)
       p. 798.

\item  V. Chari and A. Pressley,  A  Guide to Quantum Groups (Cambridge
       Univ. Press, Cambridge, 1994).

\item  K. Takemura and D. Uglov, {\it The orthogonal eigenbasis
       and norms of eigenvectors in the spin Calogero-Sutherland model },
       RIMS-1114, solv-int/9611006.

\item  B. Basu-Mallick and P. Ramadevi, Phys. Lett. A 211 (1996) 339.

\item  B. Basu-Mallick, P. Ramadevi and R. Jagannathan, {\it  Multiparametric
       and coloured extensions of the quantum group $GL_q(N)$ and the Yangian
       algebra $Y(gl_N)$ through a symmetry transformation of the Yang-Baxter
       equation},(1995) q-alg/9511028, to be published in
       Int. Jour. of Mod. Phys. A.

\item  S.M. Khoroshkin, A.A. Stolin and V.N. Tolstoy, {\it Deformation
       of Yangian $Y(sl_2)$}, TRITA-MAT-1995-MA-17, q-alg/9511005.

\item  A. Stolin and P.P. Kulish, {\it New rational solution
       of Yang-Baxter equation and deformed Yangians }, TRITA-MAT-1996-JU-11,
       q-alg/9608011.

\item  A. Kundu and B. Basu-Mallick, Mod. Phys. Lett. A 7 (1992) 61.

\item  B. Basu-Mallick and A. Kundu, Mod. Phys. Lett. A 10 (1995) 3113.

\item  A. Kundu, {\it  Quantum integrable systems: construction, solution,
       algebraic aspect, } hep-th/9612046.

\item  L.A. Takhtajan and L.D. Faddeev, Russian Math. Surveys 34 (1979) 11.

\item  L.D. Faddeev, Sov. Sc. Rev. C1 (1980) 107.

\item  P.P. Kulish and E.K. Sklyanin, Lect. Notes in Phys. (ed. J.
       Hietarinta et al, Springer, Berlin, 1982) vol. 151, p.61.

\item  C.F. Dunkl, Trans. Am. Math. Soc. 311 (1989) 167.

\item  B. Basu-Mallick, {\it Spin dependent extension of
       Calogero-Sutherland model through anyon like representations
       of permutation operators },  TIFR/TH/96-08, \hfil \break
       hep-th/9602107, to appear in Nuclear Physics B.

\item  M.L. Ge, C.P. Sun and K. Xue, Int. J. Mod. Phys. A 7 (1992) 6609.

\item  R. Chakrabarti and R. Jagannathan, Z. Phys. C 66 (1995) 523.

\item  S. Majid and M.J. Rodriguez-Plaza, J. Math. Phys. 36 (1995) 7081.

\item  A. Schirrmacher, Z. Phys. C 50 (1991) 321.

\item  J.H.H. Perk and C.L. Schultz, Phys. Lett. A 84 (1981) 407.

\item  V.F.R. Jones, Int. J. Mod. Phys. B 4 (1990) 701.

\item  V.G. Drinfeld, Soviet Math. Dokl. 27 (1983) 68; \hfil \break 
       N. Reshetikhin, Lett. Math. Phys. 20 (1990) 331.

\end{enumerate}
\end{document}